\documentclass[preprint,superscriptaddress,nofootinbib]{revtex4}
  \usepackage{graphicx}
  \begin{document}
  \title{${\Upsilon}(nS)$ ${\to}$ $B_{c}{\rho}$, $B_{c}K^{\ast}$
         decays with perturbative QCD approach}
  \author{Junfeng Sun}
  \affiliation{Institute of Particle and Nuclear Physics,
              Henan Normal University, Xinxiang 453007, China}
  \author{Yueling Yang}
  \affiliation{Institute of Particle and Nuclear Physics,
              Henan Normal University, Xinxiang 453007, China}
  \author{Jinshu Huang}
  \affiliation{College of Physics and Electronic Engineering,
              Nanyang Normal University, Nanyang 473061, China}
  \author{Gongru Lu}
  \affiliation{Institute of Particle and Nuclear Physics,
              Henan Normal University, Xinxiang 453007, China}
  \author{Qin Chang}
  \affiliation{Institute of Particle and Nuclear Physics,
              Henan Normal University, Xinxiang 453007, China}

  \begin{abstract}
  Inspired by the potential prospects of ${\Upsilon}(nS)$ data
  samples ($n$ $=$ $1$, $2$, $3$) at LHC and SuperKEKB,
  ${\Upsilon}(nS)$ ${\to}$ $B_{c}{\rho}$, $B_{c}K^{\ast}$
  decays are studied phenomenologically with pQCD approach.
  Branching ratios for ${\Upsilon}(nS)$ ${\to}$ $B_{c}{\rho}$
  and $B_{c}K^{\ast}$ decays are estimated to reach up to
  ${\cal O}(10^{-11})$ and ${\cal O}(10^{-12})$, respectively.
  Given the identification and detection efficiency
  of final states, searching for these weak decay modes should be
  fairly challenging experimentally in the future.
  \end{abstract}
  \maketitle

  \section{Introduction}
  \label{sec01}
  The spin-triplet $S$-wave $b\bar{b}$ states ${\Upsilon}(1S)$,
  ${\Upsilon}(2S)$, and ${\Upsilon}(3S)$ have some common features.
  They all lie below the open bottom threshold, and carry the
  same quantum numbers of $I^{G}J^{PC}$ $=$ $0^{-}1^{--}$ \cite{pdg}.
  For each of them, the mass is ten times as large as proton,
  but the full decay width is very narrow, only a few keV.
  Based on the above-mentioned facts, here we will use a notation
  ${\Upsilon}(nS)$ to represent special ${\Upsilon}(1S)$,
  ${\Upsilon}(2S)$, and ${\Upsilon}(3S)$ mesons for simplicity
  if it is not specified explicitly.
  Thanks to the unremitting endeavor and
  splendid performance from experimental
  groups of CLEO, CDF, D0, BaBar, Belle, LHCb, ATLAS, and so
  on, great achievements have been made in understanding
  of bottomonium properties  \cite{1212.6552}.
  The ${\Upsilon}(nS)$ decays through the strong interaction,
  electromagnetic interaction and radiative transition,
  have been extensively studied.
  The rapid accumulation of ${\Upsilon}(nS)$ data samples with
  high precision will enable a realistic possibility to search
  for ${\Upsilon}(1S)$ weak decay at the LHC and SuperKEKB.
  In this paper, we will study the ${\Upsilon}(nS)$ ${\to}$
  $B_{c}V$ weak decays ($V$ $=$ ${\rho}$, $K^{\ast}$) with
  perturbative QCD (pQCD) approach \cite{pqcd1,pqcd2,pqcd3}
  to offer a ready reference for the future experimental
  research.

  Both $b$ and $\bar{b}$ quarks in ${\Upsilon}(nS)$
  meson can decay individually via the weak interaction.
  It is well known that a clear hierarchy of the quark-mixing
  Cabibbo-Kabayashi-Maskawa (CKM) matrix elements
  opts favorably for the $b$ ${\to}$ $c$ transition, so ${\Upsilon}(nS)$
  weak decay into final states containing a $\bar{b}c$ or $b\bar{c}$
  bound state should have a relatively large branching fraction.
  Recently, we have studied the nonleptonic ${\Upsilon}(nS)$ ${\to}$
  $B_{c}^{(\ast)}P$ decays ($P$ $=$ ${\pi}$, $K$, $D$) with pQCD approach
  \cite{prd92sun,plb751,plb752,npb903,ahep4893649}, and our estimation of branching
  ratio for ${\Upsilon}(1S)$ ${\to}$ $B_{c}P$ decays is basically
  consistent with previous results using other theoretical models
  \cite{ijmpa14,ahep706543,ahep691261}.
  This positive fact encourages us to investigate other
  ${\Upsilon}(nS)$ weak decay modes.  
  The amplitudes for ${\Upsilon}(nS)$ ${\to}$ $B_{c}V$ decays
  are relatively complicated because of the $s$, $p$, $d$ wave
  contributions rather than only $p$ wave contribution
  for ${\Upsilon}(nS)$ ${\to}$ $B_{c}P$ decays.
  In addition, the ${\Upsilon}(nS)$ ${\to}$ $B_{c}V$ decays offer
  another plaza to further explore the underlying dynamical mechanism
  of heavy quarkonium weak decay.

  This paper is organized as follows.
  In section \ref{sec02}, we present
  the theoretical framework and the amplitudes for
  ${\Upsilon}(nS)$ ${\to}$ $B_{c}V$ decay.
  The numerical results and discussion are given in
  section \ref{sec03}.
  The last section is a summary.

  \section{theoretical framework}
  \label{sec02}
  \subsection{The effective Hamiltonian}
  \label{sec0201}
  Phenomenologically, assisted with the operator product expansion
  and renormalization group (RG) technique, the effective
  weak Hamiltonian accounting for ${\Upsilon}(nS)$ ${\to}$
  $B_{c}V$ decay has the following structure \cite{9512380},
   \begin{equation}
  {\cal H}_{\rm eff}\ =\ \frac{G_{F}}{\sqrt{2}}\,
   V_{cb} V_{uq}^{\ast}\,
   \Big\{ C_{1}({\mu})\,Q_{1}({\mu})
         +C_{2}({\mu})\,Q_{2}({\mu}) \Big\}
   + {\rm h.c.}
   \label{hamilton},
   \end{equation}
  where $G_{F}$ ${\simeq}$ $1.166{\times}10^{-5}\,{\rm GeV}^{-2}$
  \cite{pdg} is the Fermi constant.
  Using the Wolfenstein parameterization, the CKM factors are
  written approximately in term of $A$ and ${\lambda}$, i.e.,
  \begin{equation}
  V_{cb}V_{ud}^{\ast}\ =\
               A{\lambda}^{2}
  - \frac{1}{2}A{\lambda}^{4}
  - \frac{1}{8}A{\lambda}^{6}
  +{\cal O}({\lambda}^{8})
  \label{eq:vcbcud},
  \end{equation}
 for ${\Upsilon}(nS)$ ${\to}$ $B_{c}{\rho}$ decay, and
  \begin{equation}
  V_{cb}V_{us}^{\ast}\ =\
               A{\lambda}^{3}
  +{\cal O}({\lambda}^{8})
  \label{eq:vcbcus},
  \end{equation}
  for ${\Upsilon}(nS)$ ${\to}$ $B_{c}K^{\ast}$ decay.
  The local operators are expressed as
    \begin{eqnarray}
    Q_{1} &=&
  [ \bar{c}_{\alpha}{\gamma}_{\mu}(1-{\gamma}_{5})b_{\alpha} ]
  [ \bar{q}_{\beta} {\gamma}^{\mu}(1-{\gamma}_{5})u_{\beta} ]
    \label{q1}, \\
    Q_{2} &=&
  [ \bar{c}_{\alpha}{\gamma}_{\mu}(1-{\gamma}_{5})b_{\beta} ]
  [ \bar{q}_{\beta}{\gamma}^{\mu}(1-{\gamma}_{5})u_{\alpha} ]
    \label{q2}.
    \end{eqnarray}
  where ${\alpha}$ and ${\beta}$ are color indices,
  and $q$ denotes $d$ and $s$.

  In Eq.(\ref{hamilton}), the auxiliary scale ${\mu}$ factorizes
  physical contributions into two parts. The physical contributions above
  ${\mu}$ are integrated into the Wilson coefficients $C_{1,2}$,
  which has been reliably calculated to the next-to-leading
  order with the RG-improved perturbation theory \cite{9512380}.
  The physical contributions below ${\mu}$ are embodied in
  hadronic matrix elements (HME), where the local operators
  are sandwiched between initial and final hadron states.
  The incorporation of long distance contributions make
  HME very challenging and complicated to evaluate.
  HME is not yet fully understood so far.
  However, to obtain decay amplitudes, one has to treat
  HME with certain comprehensible approximation or assumptions,
  which result in a number of uncertainties.

  \subsection{Hadronic matrix elements}
  \label{sec0202}
  Based on factorization ansatz \cite{npb133,zpc29,npbps11} and
  hard-scattering approach \cite{plb87,prd22,plb90,prd21,plb94},
  HME has a simple structure, and is commonly expressed as a
  convolution of hard scattering kernel function ${\cal T}$
  with distribution amplitudes (DAs).
  Only DAs are nonperturbative inputs, which, on the other
  hand, are process independent, i.e., DAs determined by
  nonperturbative methods or extracted from experimental
  data can be employed to make predictions.
  With the collinear approximation, hard scattering kernels
  for annihilation contributions and spectator interactions
  can not provide sufficient endpoint suppression
  \cite{plb488,prd64,npb606}.
  In order to admit a perturbative treatment for HME,
  the intrinsic transverse momentum of valence quarks
  is kept explicitly and a Sudakov factor for each DAs
  is introduced with pQCD approach \cite{pqcd1,pqcd2,pqcd3}.
  Finally, a pQCD amplitude is written as a convolution
  integral of three parts: Wilson coefficients $C_{i}$,
  hard scattering kernel ${\cal T}$ and wave functions ${\Phi}$,
  \begin{equation}
  {\int} dk\,
  C_{i}(t)\,{\cal T}(t,k)\,{\Phi}(k)\,e^{-S}
  \label{hadronic},
  \end{equation}
  where $t$ is a typical scale, $k$ is the momentum of valence
  quarks and $e^{-S}$ is a Sudakov factor.

  \subsection{Kinematic variables}
  \label{sec0203}
\begin{figure}[ht]
  \includegraphics[width=0.99\textwidth,bb=75 620 530 725]{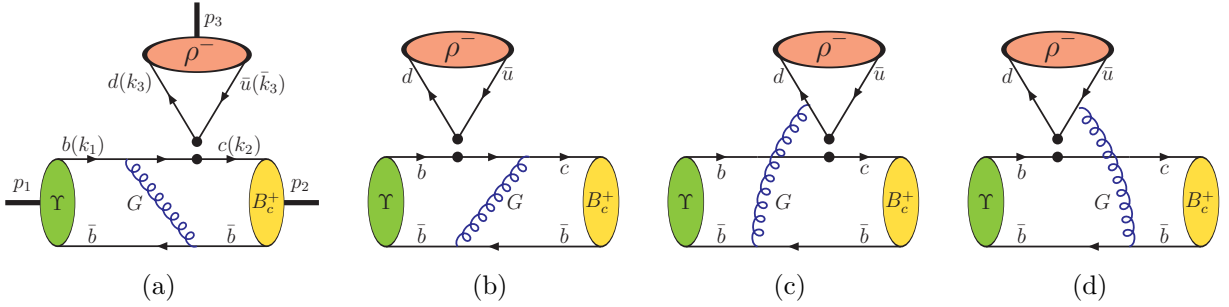}
  \caption{Feynman diagrams for ${\Upsilon}(nS)$ ${\to}$ $B_{c}{\rho}$
   decay, where (a,b) are factorizable topologies, (c,d) are
   nonfactorizable topologies.}
  \label{feynman}
  \end{figure}

  In the center-of-mass frame of ${\Upsilon}(nS)$,
  kinematic variables are defined as follows.
  \begin{equation}
  p_{{\Upsilon}}\, =\, p_{1}\, =\, \frac{m_{1}}{\sqrt{2}}(1,1,0)
  \label{kine-p1},
  \end{equation}
  \begin{equation}
  p_{B_{c}}\, =\, p_{2}\, =\, (p_{2}^{+},p_{2}^{-},0)
  \label{kine-p2},
  \end{equation}
  \begin{equation}
  p_{V}\, =\, p_{3}\, =\, (p_{3}^{-},p_{3}^{+},0)
  \label{kine-p3},
  \end{equation}
  \begin{equation}
  k_{i}\, =\, x_{i}\,p_{i}+(0,0,\vec{k}_{i{\perp}})
  \label{kine-ki},
  \end{equation}
  \begin{equation}
 {\epsilon}_{i}^{\parallel}\, =\,
  \frac{p_{i}}{m_{i}}-\frac{m_{i}}{p_{i}{\cdot}n_{+}}n_{+}
  \label{kine-longe},
  \end{equation}
  \begin{equation}
 {\epsilon}_{i}^{\perp}\, =\, (0,0,\vec{1})
  \label{kine-transe},
  \end{equation}
  \begin{equation}
  n_{+}=(1,0,0)
  \label{kine-null},
  \end{equation}
  \begin{equation}
  p_{i}^{\pm}\, =\, (E_{i}\,{\pm}\,p)/\sqrt{2}
  \label{kine-pipm},
  \end{equation}
  \begin{equation}
  s\, =\, 2\,p_{2}{\cdot}p_{3}\, =\, m_{1}^{2}-m_{2}^{2}-m_{3}^{2}
  \label{kine-s},
  \end{equation}
  \begin{equation}
  t\, =\, 2\,p_{1}{\cdot}p_{2}\, =\, m_{1}^{2}+m_{2}^{2}-m_{3}^{2}\, =\, 2\,m_{1}\,E_{2}
  \label{kine-t},
  \end{equation}
  \begin{equation}
  u\, =\, 2\,p_{1}{\cdot}p_{3}\, =\, m_{1}^{2}-m_{2}^{2}+m_{3}^{2}\, =\, 2\,m_{1}\,E_{3}
  \label{kine-u},
  \end{equation}
  \begin{equation}
  s\,t+s\,u-t\,u-4\,m_{1}^{2}\,p^{2}\, =\, 0
  \label{kine-pcm},
  \end{equation}
  where $x_{i}$ and $\vec{k}_{i{\perp}}$ are the longitudinal momentum
  fraction and transverse momentum of valence quark, respectively;
  ${\epsilon}_{i}^{\parallel}$ and ${\epsilon}_{i}^{\perp}$ are the
  longitudinal and transverse polarization vectors, respectively,
  satisfying relationship ${\epsilon}_{i}^{2}$ $=$ $-1$ and
  ${\epsilon}_{i}{\cdot}p_{i}$ $=$ $0$;
  $n_{+}$ is a positive null vector;
  the subscript $i$ $=$ $1$, $2$, $3$ on variables ($p_{i}$, $E_{i}$,
  $m_{i}$ and ${\epsilon}_{i}$) corresponds to ${\Upsilon}(nS)$,
  $B_{c}$ and $V$ mesons, respectively;
  $s$, $t$ and $u$ are Lorentz-invariant variables.
  The notation of momentum is displayed in Fig.\ref{feynman}(a).

  \subsection{Wave functions}
  \label{sec0204}
  With the notation in \cite{prd65,jhep0703},
  meson wave functions are defined as
  \begin{equation}
 {\langle}0{\vert}b_{i}(z)\bar{b}_{j}(0){\vert}
 {\Upsilon}(p_{1},{\epsilon}_{1}^{\parallel}){\rangle}\,
 =\, \frac{f_{\Upsilon}}{4}{\int}d^{4}k_{1}\,e^{-ik_{1}{\cdot}z}
  \Big\{ \!\!\not{\epsilon}_{1}^{{\parallel}} \Big[
   m_{1}\,{\Phi}_{\Upsilon}^{v}(k_{1})
  -\!\!\not{p}_{1}\, {\Phi}_{\Upsilon}^{t}(k_{1})
  \Big] \Big\}_{ji}
  \label{wave-bbl},
  \end{equation}
  \begin{equation}
 {\langle}0{\vert}b_{i}(z)\bar{b}_{j}(0){\vert}
 {\Upsilon}(p_{1},{\epsilon}_{1}^{{\perp}}){\rangle}\,
 =\, \frac{f_{\Upsilon}}{4}{\int}d^{4}k_{1}\,e^{-ik_{1}{\cdot}z}
  \Big\{ \!\!\not{\epsilon}_{1}^{{\perp}} \Big[
   m_{1}\,{\Phi}_{\Upsilon}^{V}(k_{1})
  -\!\!\not{p}_{1}\, {\Phi}_{\Upsilon}^{T}(k_{1})
  \Big] \Big\}_{ji}
  \label{wave-bbt},
  \end{equation}
  \begin{equation}
 {\langle}B_{c}(p_{2}){\vert}\bar{c}_{i}(z)b_{j}(0){\vert}0{\rangle}\,
 =\, \frac{i}{4}f_{B_{c}} {\int}dk_{2}\,e^{ik_{2}{\cdot}z}\,
  \Big\{ {\gamma}_{5}\Big[ \!\!\not{p}_{2}\,{\Phi}_{B_{c}}^{a}(k_{2})
  +m_{2}\,{\Phi}_{B_{c}}^{p}(k_{2})\Big] \Big\}_{ji}
  \label{wave-bcp},
  \end{equation}
  \begin{equation}
 {\langle}V(p_{3},{\epsilon}_{3}^{{\parallel}})
 {\vert}u_{i}(0)\bar{q}_{j}(z){\vert}0{\rangle}\, =\,
  \frac{1}{4}{\int}_{0}^{1}dk_{3}\,e^{ik_{3}{\cdot}z}
  \Big\{ \!\!\not{\epsilon}_{3}^{{\parallel}}\,
   m_{3}\,{\Phi}_{V}^{v}(k_{3})
  +\!\!\not{\epsilon}_{3}^{{\parallel}}
   \!\!\not{p}_{3}\, {\Phi}_{V}^{t}(k_{3})
  +m_{3}\,{\Phi}_{V}^{s}(k_{3}) \Big\}_{ji}
  \label{wave-vlong},
  \end{equation}
  \begin{eqnarray}
  \lefteqn{
 {\langle}V(p_{3},{\epsilon}_{3}^{{\perp}})
 {\vert}u_{i}(0)\bar{q}_{j}(z){\vert}0{\rangle}\ =\
  \frac{1}{4}{\int}_{0}^{1}dk_{3}\,e^{ik_{3}{\cdot}z}
  \Big\{ \!\not{\epsilon}_{3}^{{\perp}}
   m_{3}\,{\Phi}_{V}^{V}(k_{3})
  }
  \nonumber \\ & &
  +\!\not{\epsilon}_{3}^{{\perp}}
   \!\not{p}_{3}\, {\Phi}_{V}^{T}(k_{3})
  +\frac{i\,m_{3}}{p_{3}{\cdot}n_{+}}
  {\varepsilon}_{{\mu}{\nu}{\alpha}{\beta}}\,
  {\gamma}_{5}\,{\gamma}^{\mu}\,{\epsilon}_{3}^{{\perp},{\nu}}\,
  p_{3}^{\alpha}\,n_{+}^{\beta}\,
  {\Phi}_{V}^{A}(k_{3}) \Big\}_{ji}
  \label{wave-vtrans},
  \end{eqnarray}
  where $f_{\Upsilon}$ and $f_{B_{c}}$ are decay constants;
  ${\Phi}_{V}^{v,T}$ and ${\Phi}_{B_{c}}^{a}$ are twist-2;
  ${\Phi}_{V}^{t,s,V,A}$ and ${\Phi}_{B_{c}}^{p}$ are twist-3.

  The expressions of DAs for double heavy ${\Upsilon}(nS)$
  and $B_{c}$ mesons are \cite{plb751}
   \begin{equation}
  {\phi}_{{\Upsilon}(1S)}^{v}(x) =
  {\phi}_{{\Upsilon}(1S)}^{T}(x) = A\, x\,\bar{x}\,
  {\exp}\Big\{ -\frac{m_{b}^{2}}{8\,{\beta}_{1}^{2}\,x\,\bar{x}} \Big\}
   \label{da-bb1sa},
   \end{equation}
   \begin{equation}
  {\phi}_{{\Upsilon}(1S)}^{t}(x) = B\, (\bar{x}-x)^{2}\,
  {\exp}\Big\{ -\frac{m_{b}^{2}}{8\,{\beta}_{1}^{2}\,x\,\bar{x}} \Big\}
   \label{da-bb1sb},
   \end{equation}
   \begin{equation}
  {\phi}_{{\Upsilon}(1S)}^{V}(x) = C\, \{ 1+(\bar{x}-x)^{2} \}\,
  {\exp}\Big\{ -\frac{m_{b}^{2}}{8\,{\beta}_{1}^{2}\,x\,\bar{x}} \Big\}
   \label{da-bb1sc},
   \end{equation}
   \begin{equation}
  {\phi}_{{\Upsilon}(2S)}^{v,t,T,V}(x) = D\,
  {\phi}_{{\Upsilon}(1S)}^{v,t,T,V}(x)\,
   \Big\{ 1+\frac{m_{b}^{2}}{2\,{\beta}_{1}^{2}\,x\,\bar{x}} \Big\}
   \label{da-bb2sa},
   \end{equation}
   \begin{equation}
  {\phi}_{{\Upsilon}(3S)}^{v,t,T,V}(x) = E\,
  {\phi}_{{\Upsilon}(1S)}^{v,t,T,V}(x)\,
   \Big\{ \Big( 1-\frac{m_{b}^{2}}{2\,{\beta}_{1}^{2}\,x\,\bar{x}} \Big)^{2}
   +6 \Big\}
   \label{da-bb3sa},
   \end{equation}
   \begin{equation}
  {\phi}_{B_{c}}^{a}(x) = F\, x\,\bar{x}\,
  {\exp}\Big\{ -\frac{\bar{x}\,m_{c}^{2}+x\,m_{b}^{2}}
                     {8\,{\beta}_{2}^{2}\,x\,\bar{x}} \Big\}
   \label{da-bca},
   \end{equation}
   \begin{equation}
  {\phi}_{B_{c}}^{p}(x) = G\,
  {\exp}\Big\{ -\frac{\bar{x}\,m_{c}^{2}+x\,m_{b}^{2}}
                     {8\,{\beta}_{2}^{2}\,x\,\bar{x}} \Big\}
   \label{da-bcp},
   \end{equation}
   where $\bar{x}$ $=$ $1$ $-$ $x$;
   ${\beta}_{i}$ ${\simeq}$ $m_{i}\,{\alpha}_{s}(m_{i})$
   according to nonrelativistic quantum chromodynamics
   (NRQCD) power counting rules \cite{prd46,prd51,rmp77};
   parameters $A$, $B$, $C$, $D$, $E$, $F$, $G$ are
   normalization coefficients satisfying the conditions
   \begin{equation}
  {\int}_{0}^{1}dx\,{\phi}_{\Upsilon}^{v,t,V,T}(x) =1,
   \qquad
  {\int}_{0}^{1}dx\,{\phi}_{B_{c}}^{a,p}(x)=1
   \label{wave-abc}.
   \end{equation}

\begin{figure}[ht]
  \includegraphics[width=0.8\textwidth,bb=75 430 530 710]{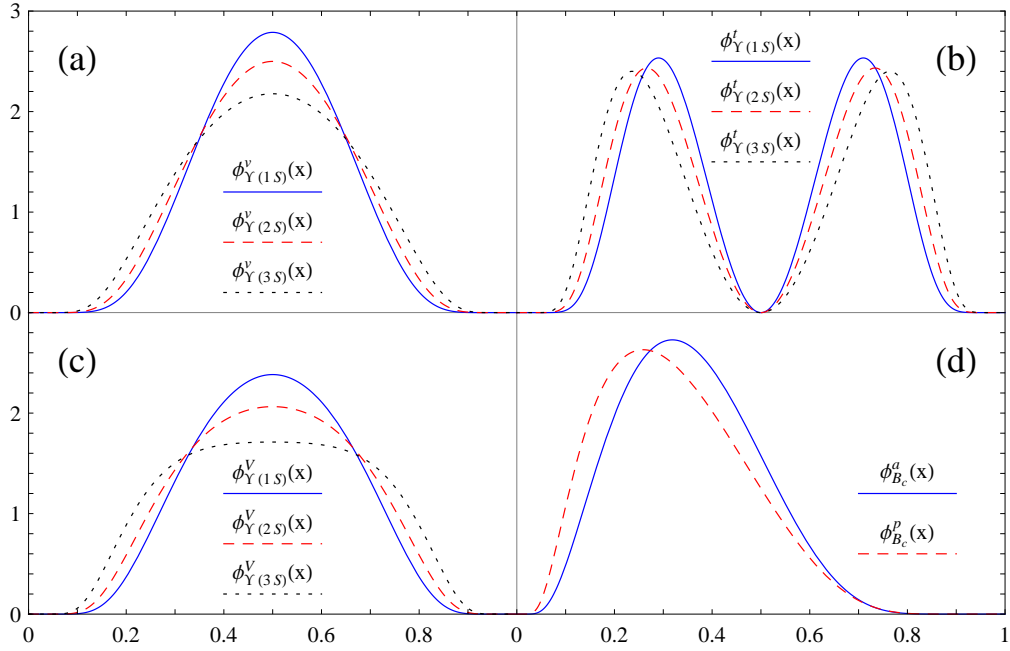}
  \caption{The normalized distribution amplitudes for
   ${\Upsilon}(nS)$ and $B_{c}$ mesons.}
  \label{fig:wave}
  \end{figure}

  The shape lines of DAs for ${\Upsilon}(nS)$ and $B_{c}$
  mesons are showed in Fig. \ref{fig:wave}.
  It is clearly seen that
  (1) DAs for ${\Upsilon}(nS)$ and $B_{c}$ are basically consistent
  with a picture that valence quarks share momentum fractions
  according to their masses;
  (2) DAs fall quickly down to zero at endpoint $x$, $\bar{x}$
  ${\to}$ $0$ due to suppression from exponential functions,
  which are bound to offer a natural and effective cutoff for
  soft contributions.

  For the light vector mesons, only three wave functions
  ${\Phi}_{V}^{v}$ and ${\Phi}_{V}^{V,A}$ are involved
  in actual calculation (see Appendix).
  Their asymptotic forms are \cite{prd65,jhep0703}:
  \begin{equation}
 {\phi}_{V}^{v}(x) \, =\, 6\,x\,\bar{x}
  \label{da-rhov},
  \end{equation}
  \begin{equation}
 {\phi}_{V}^{V}(x) \, =\, \frac{3}{4}\, \Big\{ 1+(\bar{x}-x)^{2} \Big\}
  \label{da-rhoV},
  \end{equation}
  \begin{equation}
 {\phi}_{V}^{A}(x) \, =\, \frac{3}{2}\, (\bar{x}-x)
  \label{da-rhoA}.
  \end{equation}

  \subsection{Decay amplitudes}
  \label{sec0205}
  The Feynman diagrams for ${\Upsilon}(nS)$ ${\to}$
  $B_{c}{\rho}$ decay are shown in Fig.\ref{feynman},
  including factorizable emission topologies (a) and (b)
  where gluon connects initial ${\Upsilon}(nS)$ with
  recoiled $B_{c}$ mesons,
  and nonfactorizable emission topologies (c) and (d)
  where gluon attaches the spectator quark with
  emitted vector mesons.

  After a straightforward calculation,
  amplitude for ${\Upsilon}(nS)$ ${\to}$ $B_{c}V$
  decay can be decomposed as below,
   \begin{equation}
  {\cal A}({\Upsilon}(nS){\to}B_{c}V)\ =\
  {\cal A}_{L}({\epsilon}_{1}^{{\parallel}},{\epsilon}_{3}^{{\parallel}})
 +{\cal A}_{N}({\epsilon}_{1}^{{\perp}}{\cdot}{\epsilon}_{3}^{{\perp}})
 +i\,{\cal A}_{T}\,{\varepsilon}_{{\mu}{\nu}{\alpha}{\beta}}\,
  {\epsilon}_{1}^{{\mu}}\,{\epsilon}_{3}^{{\nu}}\,
   p_{1}^{\alpha}\,p_{3}^{\beta}
   \label{eq:amp01},
   \end{equation}
  which is conventionally written as helicity amplitudes,
   \begin{equation}
  {\cal A}_{0}\ =\ -{\cal C}\,\sum\limits_{i}
  {\cal A}_{i,L}({\epsilon}_{1}^{{\parallel}},{\epsilon}_{3}^{{\parallel}})
   \label{eq:amp02},
   \end{equation}
   \begin{equation}
  {\cal A}_{\parallel}\ =\ \sqrt{2}\,{\cal C} \sum\limits_{i}
  {\cal A}_{i,N}
   \label{eq:amp03},
   \end{equation}
   \begin{equation}
  {\cal A}_{\perp}\ =\ \sqrt{2}\,{\cal C}\,m_{1}\,p \sum\limits_{i}
  {\cal A}_{i,T}
   \label{eq:amp04},
   \end{equation}
   \begin{equation}
  {\cal C}\ =\ i\frac{G_{F}}{\sqrt{2}}\,\frac{{\pi}\,C_{F}}{N_{c}}\,
   f_{\Upsilon}\,f_{B_{c}}\, f_{V}\, V_{cb} V_{uq}^{\ast}
   \label{eq:amp05},
   \end{equation}
  where $C_{F}$ $=$ $4/3$ and the color number $N_{c}$ $=$ $3$;
  the first superscript $i$ on ${\cal A}_{i,L(N,T)}$ corresponds
  to the indices of Fig.\ref{feynman}.
  The detailed analytical expressions of building blocks
  ${\cal A}_{i,L(N,T)}$ are displayed in Appendix.

  \section{Numerical results and discussion}
  \label{sec03}

  In the rest frame of ${\Upsilon}(nS)$, decaying into
  $B_{c}$ and light vector $V$ mesons,
  branching ratio is defined as
   \begin{equation}
  {\cal B}r\ =\ \frac{1}{12{\pi}}\,
   \frac{p}{m_{{\Upsilon}}^{2}{\Gamma}_{{\Upsilon}}}\, \Big\{
  {\vert}{\cal A}_{0}{\vert}^{2}+{\vert}{\cal A}_{\parallel}{\vert}^{2}
 +{\vert}{\cal A}_{\perp}{\vert}^{2} \Big\}
   \label{br},
   \end{equation}
  where $p$ is the center-of-mass momentum of final states.

   \begin{table}[ht]
   \caption{The numerical values of input parameters.}
   \label{tab:input}
   \begin{ruledtabular}
   \begin{tabular}{lll}
   \multicolumn{3}{c}{CKM parameter \cite{pdg}} \\ \hline
   \multicolumn{3}{c}{
    $A$          $=$ $0.814^{+0.023}_{-0.024}$, \qquad
    ${\lambda}$  $=$ $0.22537{\pm}0.00061$, } \\ \hline
    \multicolumn{3}{c}{mass, width and decay constant} \\ \hline
    $m_{{\Upsilon}(1S)}$ $=$ $9460.30{\pm}0.26$ MeV \cite{pdg},
  & ${\Gamma}_{{\Upsilon}(1S)}$ $=$ $54.02{\pm}1.25$ keV \cite{pdg},
  & $f_{{\Upsilon}(1S)}$ $=$ $676.4{\pm}10.7$ MeV \cite{plb751}, \\
    $m_{{\Upsilon}(2S)}$ $=$ $10023.26{\pm}0.31$ MeV \cite{pdg},
  & ${\Gamma}_{{\Upsilon}(2S)}$ $=$ $31.98{\pm}2.63$ keV \cite{pdg},
  & $f_{{\Upsilon}(2S)}$ $=$ $473.0{\pm}23.7$ MeV \cite{plb751}, \\
    $m_{{\Upsilon}(3S)}$ $=$ $10355.2{\pm}0.5$ MeV \cite{pdg},
  & ${\Gamma}_{{\Upsilon}(3S)}$ $=$ $20.32{\pm}1.85$ keV \cite{pdg},
  & $f_{{\Upsilon}(3S)}$ $=$ $409.5{\pm}29.4$ MeV \cite{plb751}, \\
    $m_{B_{c}}$ $=$ $6275.6{\pm}1.1$ MeV \cite{pdg},
  & $m_{b}$ $=$ $4.78{\pm}0.06$ GeV \cite{pdg},
  & $m_{c}$ $=$ $1.67{\pm}0.07$ GeV \cite{pdg}, \\
    $f_{B_{c}}$ $=$ $434{\pm}15$ MeV \cite{prd91},
  & $f_{\rho}$ $=$ $216{\pm}3$ MeV \cite{jhep0703},
  & $f_{K^{\ast}}$ $=$ $220{\pm}5$ MeV \cite{jhep0703}
  \end{tabular}
  \end{ruledtabular}
  \end{table}
   \begin{table}[ht]
   \caption{Branching ratio for ${\Upsilon}(nS)$ ${\to}$
   $B_{c}{\rho}$, $B_{c}K^{\ast}$.}
   \label{tab:br}
   \begin{ruledtabular}
   \begin{tabular}{lcccc}
  & this work
  & Ref. \cite{ijmpa14}
  & Ref. \cite{ahep706543}
  & Ref. \cite{ahep691261} \\ \hline
    $10^{11}{\times}{\cal B}r({\Upsilon}(1S){\to}B_{c}{\rho})$
  & $13.25^{+1.04+1.14+0.91}_{-0.63-1.14-0.87}$
  & $17.6$
  & $13.0$
  & $15.3$ \\ \hline
    $10^{11}{\times}{\cal B}r({\Upsilon}(2S){\to}B_{c}{\rho})$
  & $8.88^{+0.64+0.67+0.61}_{-0.40-0.74-0.58}$
  & ... & ... & ... \\ \hline
    $10^{11}{\times}{\cal B}r({\Upsilon}(3S){\to}B_{c}{\rho})$
  & $8.46^{+0.61+0.71+0.58}_{-0.37-0.68-0.56}$
  & ... & ... & ... \\ \hline
    $10^{12}{\times}{\cal B}r({\Upsilon}(1S){\to}B_{c}K^{\ast})$
  & $7.97^{+0.62+0.65+0.59}_{-0.38-0.67-0.57}$
  & $10.0$
  & $7.0$
  & $8.75$ \\ \hline
    $10^{12}{\times}{\cal B}r({\Upsilon}(2S){\to}B_{c}K^{\ast})$
  & $5.28^{+0.38+0.48+0.39}_{-0.24-0.45-0.37}$
  & ... & ... & ... \\ \hline
    $10^{12}{\times}{\cal B}r({\Upsilon}(3S){\to}B_{c}K^{\ast})$
  & $4.98^{+0.36+0.44+0.37}_{-0.22-0.43-0.35}$
  & ... & ... & ...
  \end{tabular}
  \end{ruledtabular}
  \end{table}

  The values of input parameters are listed in Table \ref{tab:input}.
  If it is not specified explicitly, their central values will be
  used as default inputs.
  Our numerical results are presented in Table \ref{tab:br},
  where the uncertainties come from scale $(1{\pm}0.1)t_{i}$,
  $m_{b}$ and $m_{c}$, and CKM parameters, respectively.
  The following are some comments.

  (1)
  By and large, our results are consistent with previous estimation
  on branching ratio for ${\Upsilon}(1S)$ ${\to}$ $B_{c}{\rho}$,
  $B_{c}K^{\ast}$ decays.
  The hierarchical structure of CKM factors
  ${\vert}V_{cb}V_{ud}^{\ast}{\vert}$ $>$ ${\vert}V_{cb}V_{us}^{\ast}{\vert}$
  leads to the general rank-size relationship among branching ratios
  ${\cal B}r({\Upsilon}(nS){\to}B_{c}{\rho})$ $>$
  ${\cal B}r({\Upsilon}(nS){\to}B_{c}K^{\ast})$.
  Normally, there should be
  ${\cal B}r({\Upsilon}(3S){\to}B_{c}V)$ $>$
  ${\cal B}r({\Upsilon}(2S){\to}B_{c}V)$ $>$
  ${\cal B}r({\Upsilon}(1S){\to}B_{c}V)$ for the
  same final $V$ meson, due to the
  fact that $m_{{\Upsilon}(3S)}$ $>$ $m_{{\Upsilon}(2S)}$
  $>$ $m_{{\Upsilon}(1S)}$ and ${\Gamma}_{{\Upsilon}(3S)}$
  $<$ ${\Gamma}_{{\Upsilon}(2S)}$
  $<$ ${\Gamma}_{{\Upsilon}(1S)}$.
  However, the numbers in Table \ref{tab:br} are beyond
  expectation. Why is it that? In addition to form factors,
  one of the possible factors is
   \begin{eqnarray} & &
  {\cal B}r({\Upsilon}(3S){\to}B_{c}V) :
  {\cal B}r({\Upsilon}(2S){\to}B_{c}V) :
  {\cal B}r({\Upsilon}(1S){\to}B_{c}V)
   \nonumber \\ &{\propto}&
   \frac{f^{2}_{{\Upsilon}(3S)}}{m^{2}_{{\Upsilon}(3S)}\,{\Gamma}_{{\Upsilon}(3S)}} :
   \frac{f^{2}_{{\Upsilon}(2S)}}{m^{2}_{{\Upsilon}(2S)}\,{\Gamma}_{{\Upsilon}(2S)}} :
   \frac{f^{2}_{{\Upsilon}(1S)}}{m^{2}_{{\Upsilon}(1S)}\,{\Gamma}_{{\Upsilon}(1S)}}
   \ {\simeq}\ 0.8 : 0.7 : 1
   \label{eq:bf-phase}.
   \end{eqnarray}

  (2)
  Branching ratio for ${\Upsilon}(nS)$ ${\to}$ $B_{c}{\rho}$
  decay can reach up to ${\cal O}(10^{-11})$.
  The ${\Upsilon}(nS)$ production cross section in
  p-Pb collision is about a few ${\mu}b$ at LHCb \cite{jhep1407}
  and ALICE \cite{plb740}. Over $10^{11}$ ${\Upsilon}(nS)$
  data samples per $ab^{-1}$ data collected at LHCb and ALICE
  are in principle available, corresponding to dozens of
  ${\Upsilon}(nS)$ ${\to}$ $B_{c}{\rho}$ events.
  If the experimental identification of final states is considered,
  for example, the best experimental identification of $B_{c}$
  meson is through $B_{c}$ ${\to}$ $J/{\psi}{\mu}^{+}{\nu}_{\mu}$
  or $J/{\psi}{\pi}$ decays with branching ratios ${\cal O}(10^{-3})$
  ${\sim}$ ${\cal O}(10^{-4})$ \cite{prd77,epjc60,prd90}
  and detection efficiency about ${\cal O}(10^{-2})$ \cite{prd90,prd92},
  then the feasibility of observation of ${\Upsilon}(nS)$ ${\to}$ $B_{c}V$
  decays is very small.

  \begin{figure}[ht]
  \includegraphics[width=0.98\textwidth,bb=80 630 530 720]{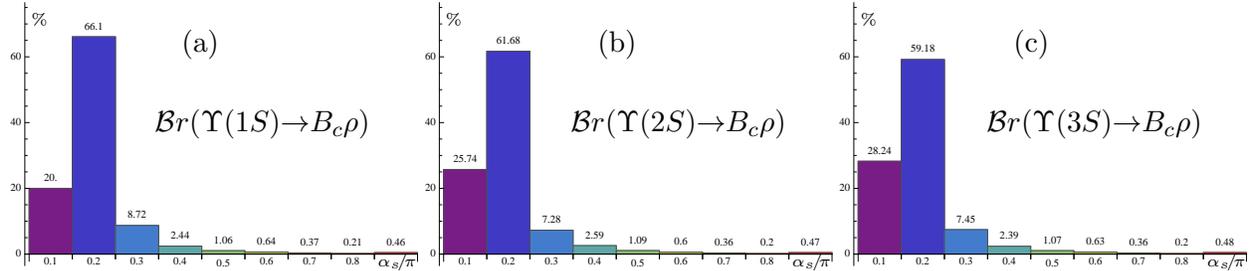}
  \caption{Contributions to branching ratio from different region
  of ${\alpha}_{s}/{\pi}$ (abscissa axis), where the numbers over
  histogram denote percentage of the corresponding contributions.}
  \label{fig:his}
  \end{figure}

 (3)
  From Fig.\ref{feynman}, the spectator is a
  heavy bottom quark in the ${\Upsilon}(nS)$ ${\to}$ $B_{c}$
  transition.
  It is assumed that the bottom quark is near on-shell
  and the gluon attaching to the spectator might be soft.
  It is natural to question the validity of perturbative
  calculation with pQCD approach.
  So, it is necessary to check how many shares come from
  the perturbative region.
  The contributions to branching ratio
  ${\cal B}r({\Upsilon}(nS){\to}B_{c}{\rho})$ from different region
  of ${\alpha}_{s}/{\pi}$ are displayed in Fig.\ref{fig:his}.
  It is clearly seen that more than 85\% (some 95\%) contributions
  to branching ratio come from ${\alpha}_{s}/{\pi}$ ${\le}$
  $0.2$ ($0.3$) regions, which implies that the calculation
  with pQCD approach is feasible.
  Compared with contributions from ${\alpha}_{s}/{\pi}$ ${\in}$
  $[0.1,0.2]$ region, one of crucial reasons for a small percentage in the region
  ${\alpha}_{s}/{\pi}$ ${\le}$ $0.1$ is that the absolute values
  of Wilson coefficients $C_{1,2}$, parameter $a_{1}$ and coupling
  ${\alpha}_{s}$ decrease along with the increase of renormalization
  scale.

  (4)
  Besides uncertainties listed in Table \ref{tab:br},
  decay constants $f_{\Upsilon}$ and $f_{B_{c}}$ can
  bring some 8\%, 12\%, 16\% uncertainties for ${\Upsilon}(1S)$,
  ${\Upsilon}(2S)$, ${\Upsilon}(3S)$ decays, respectively.
  These are two ways to reduce theoretical uncertainty.
  One is to construct some relative ratios of branching ratios,
  for example, ${\cal B}r({\Upsilon}(nS){\to}B_{c}K^{\ast})/{\cal B}r({\Upsilon}(nS){\to}B_{c}{\rho})$
  and ${\cal B}r({\Upsilon}(mS){\to}B_{c}{\rho})/{\cal B}r({\Upsilon}(nS){\to}B_{c}{\rho})$.
  The other is to consider higher order corrections to HME,
  relativistic effects on DAs, and so on. Here,
  our results just provide an order of magnitude estimation.

  \section{Summary}
  \label{sec04}
  Besides the predominant strong and electromagnetic decay modes,
  ${\Upsilon}(nS)$ can also decay through the weak interaction
  within the standard model.
  Study of ${\Upsilon}(nS)$ weak decay is theoretically
  interesting and experimentally feasible.
  In this paper, we investigated the bottom- and charm-changing
  ${\Upsilon}(nS)$ ${\to}$ $B_{c}{\rho}$, $B_{c}K^{\ast}$
  decays with phenomenological pQCD approach.
  It is found that branching ratio for ${\Upsilon}(nS)$
  ${\to}$ $B_{c}{\rho}$ and $B_{c}K^{\ast}$ decays can reach up to
  ${\cal O}(10^{-11})$ and ${\cal O}(10^{-12})$, respectively,
  and their actual detection at the future LHC and SuperKEKB experiments
  should be quite challenging.

  \section*{Acknowledgments}
  The work is supported by National Natural Science Foundation
  of China (Grant Nos. 11547014, 11475055, U1332103 and 11275057).
  We thank the referees for their constructive comments.

  \begin{appendix}
  \section{Building blocks of decay amplitudes}
  \label{blocks}
  The amplitude for the ${\Upsilon}(nS)$ ${\to}$ $B_{c}V$ decays
  ($V$ $=$ ${\rho}$, $K^{\ast}$) are constituted of a linear
  combination of building block ${\cal A}_{i,j}$, where
  the first subscript $i$ corresponds to the indices of
  Fig.\ref{feynman}, and the second subscript $j$ $=$ $L$,
  $N$, $T$ denotes to three different helicity amplitudes.
  The expressions of ${\cal A}_{i,j}$ are written as follows.
   \begin{eqnarray}
  {\cal A}_{a,L} &=&
  {\int}_{0}^{1}dx_{1}
  {\int}_{0}^{1}dx_{2}
  {\int}_{0}^{\infty}b_{1}db_{1}
  {\int}_{0}^{\infty}b_{2}db_{2}\,
  H_{f}({\alpha}_{g},{\beta}_{a},b_{1},b_{2})\,
  E_{f}(t_{a})\, {\alpha}_{s}(t_{a})\, a_{1}(t_{a})
   \nonumber \\ &{\times}&
  {\phi}_{\Upsilon}^{v}(x_{1})\, \Big\{
  {\phi}_{B_{c}}^{a}(x_{2})\,
   \Big[ m_{1}^{2}\,s -(4\,m_{1}^{2}\,p^{2}
  +m_{2}^{2}\,u)\,\bar{x}_{2} \Big]
  +{\phi}_{B_{c}}^{p}(x_{2})\,
   m_{2}\,m_{b}\,u \Big\}
   \label{amp:al}, \\
  {\cal A}_{a,N} &=&
  m_{1}\, m_{3}
  {\int}_{0}^{1}dx_{1}
  {\int}_{0}^{1}dx_{2}
  {\int}_{0}^{\infty}b_{1}db_{1}
  {\int}_{0}^{\infty}b_{2}db_{2}\,
  H_{f}({\alpha}_{g},{\beta}_{a},b_{1},b_{2})\,
  E_{f}(t_{a})\, {\alpha}_{s}(t_{a})
   \nonumber \\ &{\times}&
  a_{1}(t_{a})\,
  {\phi}_{\Upsilon}^{V}(x_{1})\, \Big\{
  {\phi}_{B_{c}}^{a}(x_{2})\, (2\,m_{2}^{2}\,\bar{x}_{2}-t)
 -{\phi}_{B_{c}}^{p}(x_{2})\,2\,m_{2}\,m_{b} \Big\}
   \label{amp:an}, \\
  {\cal A}_{a,T} &=&
  2\, m_{1}\,m_{3}
  {\int}_{0}^{1}dx_{1}
  {\int}_{0}^{1}dx_{2}
  {\int}_{0}^{\infty}b_{1}db_{1}
  {\int}_{0}^{\infty}b_{2}db_{2}\,
  H_{f}({\alpha}_{g},{\beta}_{a},b_{1},b_{2})\,
  E_{f}(t_{a})
   \nonumber \\ &{\times}&
  {\alpha}_{s}(t_{a})\,a_{1}(t_{a})\,
  {\phi}_{\Upsilon}^{V}(x_{1})\,
  {\phi}_{B_{c}}^{a}(x_{2})
   \label{amp:at},
   \end{eqnarray}
   \begin{eqnarray}
  {\cal A}_{b,L} &=&
  {\int}_{0}^{1}dx_{1}
  {\int}_{0}^{1}dx_{2}
  {\int}_{0}^{\infty}b_{1}db_{1}
  {\int}_{0}^{\infty}b_{2}db_{2}\,
  H_{f}({\alpha}_{g},{\beta}_{b},b_{2},b_{1})\,
  E_{f}(t_{b})\,
  {\alpha}_{s}(t_{b})\,
  a_{1}(t_{b})
   \nonumber \\ &{\times}&
   \Big\{ {\phi}_{\Upsilon}^{v}(x_{1})\,
   \Big[ {\phi}_{B_{c}}^{a}(x_{2})\,
   \{ m_{1}^{2}\, (s-4\,p^{2})\,\bar{x}_{1}
   -m_{2}^{2}\,u \}
 +{\phi}_{B_{c}}^{p}(x_{2})\, 2\,m_{2}\,m_{c}\,u \Big]
   \nonumber \\ & & +\,
  {\phi}_{\Upsilon}^{t}(x_{1})\,
   \Big[ {\phi}_{B_{c}}^{p}(x_{2})\, 2\,m_{1}\,m_{2}\,
   (s-u\,\bar{x}_{1})
  - {\phi}_{B_{c}}^{a}(x_{2})\,m_{1}\,m_{c}\,s \Big] \Big\}
   \label{amp:bl}, \\
  {\cal A}_{b,N} &=&
  m_{3}
  {\int}_{0}^{1}dx_{1}
  {\int}_{0}^{1}dx_{2}
  {\int}_{0}^{\infty}b_{1}db_{1}
  {\int}_{0}^{\infty}b_{2}db_{2}\,
  H_{f}({\alpha}_{g},{\beta}_{b},b_{2},b_{1})\,
  E_{f}(t_{b})\,
  {\alpha}_{s}(t_{b})
   \nonumber \\ &{\times}&
  a_{1}(t_{b})
   \Big\{ {\phi}_{\Upsilon}^{V}(x_{1})\,m_{1}\,
   \Big[ {\phi}_{B_{c}}^{a}(x_{2})\, (2\,m_{2}^{2} -t\,\bar{x}_{1})
  -{\phi}_{B_{c}}^{p}(x_{2})\,4\,m_{2}\,m_{c} \Big]
   \nonumber \\ & & +\,
  {\phi}_{\Upsilon}^{T}(x_{1})\, \Big[
  {\phi}_{B_{c}}^{a}(x_{2})\,m_{c}\,t
 +{\phi}_{B_{c}}^{p}(x_{2})\, m_{2}\,(4\,m_{1}^{2}\,\bar{x}_{1}
  -2\,t) \Big] \Big\}
   \label{amp:bn}, \\
  {\cal A}_{b,T} &=&
  -2\,m_{3}\,
  {\int}_{0}^{1}dx_{1}
  {\int}_{0}^{1}dx_{2}
  {\int}_{0}^{\infty}b_{1}db_{1}
  {\int}_{0}^{\infty}b_{2}db_{2}\,
  H_{f}({\alpha}_{g},{\beta}_{b},b_{2},b_{1})\,
  E_{f}(t_{b})\,{\alpha}_{s}(t_{b})
   \nonumber \\ &{\times}&
  a_{1}(t_{b})
   \Big\{ {\phi}_{\Upsilon}^{V}(x_{1})\,
  {\phi}_{B_{c}}^{a}(x_{2})\, m_{1}\,\bar{x}_{1}
 +{\phi}_{\Upsilon}^{T}(x_{1})\,
   \Big[ {\phi}_{B_{c}}^{a}(x_{2})\,m_{c}
 -{\phi}_{B_{c}}^{p}(x_{2})\,2\,m_{2} \Big] \Big\}
   \label{amp:bt},
   \end{eqnarray}
   \begin{eqnarray}
  {\cal A}_{c,L} &=&
   \frac{1}{N_{c}}
  {\int}_{0}^{1}dx_{1}
  {\int}_{0}^{1}dx_{2}
  {\int}_{0}^{1}dx_{3}
  {\int}_{0}^{\infty}db_{1}
  {\int}_{0}^{\infty}b_{2}db_{2}
  {\int}_{0}^{\infty}b_{3}db_{3}\,
  H_{n}({\alpha}_{g},{\beta}_{c},b_{2},b_{3})
   \nonumber \\ &{\times}&
  {\delta}(b_{1}-b_{2})\,
  E_{n}(t_{c})\,
  {\alpha}_{s}(t_{c})\,
   \Big\{ {\phi}_{\Upsilon}^{v}(x_{1})\,
  {\phi}_{B_{c}}^{a}(x_{2})\, u\,
   ( t\,x_{1}-2\,m_{2}^{2}\,x_{2}-s\,\bar{x}_{3} )
   \nonumber \\ & & +\,
  {\phi}_{\Upsilon}^{t}(x_{1})\,
  {\phi}_{B_{c}}^{p}(x_{2})\, m_{1}\,m_{2}\,
   ( s\,x_{2}+2\,m_{3}^{2}\,\bar{x}_{3}-u\,x_{1}) \Big\}\,
  {\phi}_{V}^{v}(x_{3})\, C_{2}(t_{c})
   \label{amp:cl}, \\
  {\cal A}_{c,N} &=&
   \frac{ m_{3} }{N_{c}}
  {\int}_{0}^{1}dx_{1}
  {\int}_{0}^{1}dx_{2}
  {\int}_{0}^{1}dx_{3}
  {\int}_{0}^{\infty}db_{1}
  {\int}_{0}^{\infty}b_{2}db_{2}
  {\int}_{0}^{\infty}b_{3}db_{3}\,
  H_{n}({\alpha}_{g},{\beta}_{c},b_{2},b_{3})
   \nonumber \\ &{\times}&
  {\delta}(b_{1}-b_{2})\,
   \Big\{ {\phi}_{\Upsilon}^{V}(x_{1})\,
  {\phi}_{B_{c}}^{a}(x_{2})\,
  {\phi}_{V}^{V}(x_{3})\,m_{1}\,
   ( 2\,s\,\bar{x}_{3}+4\,m_{2}^{2}\,x_{2}\,-2\,t\,x_{1} )
   \nonumber \\ & &  +\,
   {\phi}_{\Upsilon}^{T}(x_{1})\,
   {\phi}_{B_{c}}^{p}(x_{2})\, m_{2}\, \Big[
   {\phi}_{V}^{V}(x_{3})\,
   ( 2\,m_{1}^{2}\,x_{1} -t\,x_{2}-u\,\bar{x}_{3} )
   \nonumber \\ & &  +\,
  {\phi}_{V}^{A}(x_{3})\,
   2\,m_{1}\,p\,( x_{2}-\bar{x}_{3} )
   \Big] \Big\}\,
   E_{n}(t_{c})\,
   {\alpha}_{s}(t_{c})\,
   C_{2}(t_{c})
   \label{amp:cn}, \\
  {\cal A}_{c,T} &=&
   \frac{ m_{3} }{N_{c}\,p}
  {\int}_{0}^{1}dx_{1}
  {\int}_{0}^{1}dx_{2}
  {\int}_{0}^{1}dx_{3}
  {\int}_{0}^{\infty}db_{1}
  {\int}_{0}^{\infty}b_{2}db_{2}
  {\int}_{0}^{\infty}b_{3}db_{3}\,
  H_{n}({\alpha}_{g},{\beta}_{c},b_{2},b_{3})
   \nonumber \\ &{\times}&
  {\delta}(b_{1}-b_{2})\,
   \Big\{ {\phi}_{\Upsilon}^{V}(x_{1})\,
  {\phi}_{B_{c}}^{a}(x_{2})\,
  {\phi}_{V}^{A}(x_{3})\,
   ( 2\,s\,\bar{x}_{3}+4\,m_{2}^{2}\,x_{2}\,-2\,t\,x_{1} )
   \nonumber \\ & & +\,
   {\phi}_{\Upsilon}^{T}(x_{1})\,
   {\phi}_{B_{c}}^{p}(x_{2})\, \Big[
   {\phi}_{V}^{A}(x_{3})\,m_{2}/m_{1}\,
   ( 2\,m_{1}^{2}\,x_{1} -t\,x_{2}-u\,\bar{x}_{3} )
    \nonumber \\ & &  +\,
   {\phi}_{V}^{V}(x_{3})\, 2\,m_{2}\,p\,
   (x_{2}-\bar{x}_{3}) \Big] \Big\}\,
   E_{n}(t_{c})\,
   {\alpha}_{s}(t_{c})\,
   C_{2}(t_{c})
   \label{amp:ct},
   \end{eqnarray}
   \begin{eqnarray}
  {\cal A}_{d,L} &=&
   \frac{1}{N_{c}}
  {\int}_{0}^{1}dx_{1}
  {\int}_{0}^{1}dx_{2}
  {\int}_{0}^{1}dx_{3}
  {\int}_{0}^{\infty}db_{1}
  {\int}_{0}^{\infty}b_{2}db_{2}
  {\int}_{0}^{\infty}b_{3}db_{3}\,
   H_{n}({\alpha}_{g},{\beta}_{d},b_{2},b_{3})
   \nonumber \\ &{\times}&
  {\delta}(b_{1}-b_{2})\,
  {\phi}_{V}^{v}(x_{3})\, \Big\{
  {\phi}_{\Upsilon}^{t}(x_{1})\,
  {\phi}_{B_{c}}^{p}(x_{2})\, m_{1}\,m_{2}\,
  (s\,x_{2}+2\,m_{3}^{2}\,x_{3}-u\,x_{1})
   \nonumber \\ & & +\,
  {\phi}_{\Upsilon}^{v}(x_{1})\,
  {\phi}_{B_{c}}^{a}(x_{2})\,
  4\,m_{1}^{2}\,p^{2}\,
  (x_{3}-x_{2}) \Big\}\,
   E_{n}(t_{d})\,
  {\alpha}_{s}(t_{d})\,
  C_{2}(t_{d})
   \label{amp:dl}, \\
  {\cal A}_{d,N} &=&
   \frac{ m_{2}\,m_{3} }{N_{c}}
  {\int}_{0}^{1}dx_{1}
  {\int}_{0}^{1}dx_{2}
  {\int}_{0}^{1}dx_{3}
  {\int}_{0}^{\infty}db_{1}
  {\int}_{0}^{\infty}b_{2}db_{2}
  {\int}_{0}^{\infty}b_{3}db_{3}\,
  {\delta}(b_{1}-b_{2})
   \nonumber \\ &{\times}&
  H_{n}({\alpha}_{g},{\beta}_{d},b_{2},b_{3})\,
  E_{n}(t_{d})\, {\alpha}_{s}(t_{d})\,
   \Big\{ {\phi}_{V}^{V}(x_{3})\,
   ( 2\,m_{1}^{2}\,x_{1}-t\,x_{2}\,-u\,x_{3} )
   \nonumber \\ & & +\,
  {\phi}_{V}^{A}(x_{3})\, 2\,m_{1}\,p\,
  (x_{2}-x_{3}) \Big\}\,
  {\phi}_{\Upsilon}^{T}(x_{1})\,
  {\phi}_{B_{c}}^{p}(x_{2})\,
   C_{2}(t_{d})
   \label{amp:dn}, \\
  {\cal A}_{d,T} &=&
   \frac{ m_{2}\,m_{3} }{N_{c}\,m_{1}\,p}
  {\int}_{0}^{1}dx_{1}
  {\int}_{0}^{1}dx_{2}
  {\int}_{0}^{1}dx_{3}
  {\int}_{0}^{\infty}db_{1}
  {\int}_{0}^{\infty}b_{2}db_{2}
  {\int}_{0}^{\infty}b_{3}db_{3}\,
  {\delta}(b_{1}-b_{2})
   \nonumber \\ &{\times}&
  H_{n}({\alpha}_{g},{\beta}_{d},b_{2},b_{3})\,
  E_{n}(t_{d})\, {\alpha}_{s}(t_{d})\,
   \Big\{ {\phi}_{V}^{A}(x_{3})\,
   ( 2\,m_{1}^{2}\,x_{1}-t\,x_{2}\,-u\,x_{3} )
   \nonumber \\ & & +\,
  {\phi}_{V}^{V}(x_{3})\, 2\,m_{1}\,p\,
  (x_{2}-x_{3}) \Big\}\,
  {\phi}_{\Upsilon}^{T}(x_{1})\,
  {\phi}_{B_{c}}^{p}(x_{2})\,
   C_{2}(t_{d})
   \label{amp:dt},
   \end{eqnarray}
  where $x_{i}$ and $\bar{x}_{i}$ $=$ $1$ $-$ $x_{i}$
  are longitudinal momentum fractions of valence quarks;
  $b_{i}$ is the conjugate variable of the transverse
  momentum $k_{i{\perp}}$;
  $a_{1}$ $=$ $C_{1}$ $+$ $C_{2}/N_{c}$;
  $N_{c}$ $=$ $3$ is the color number;
  $C_{1,2}$ are the Wilson coefficients.

  The Sudakov factor $E_{f,n}$ and function $H_{f,n}$ are defined
  as follows, where the subscript $f$ ($n$) corresponds to
  (non)factorizable topologies.
   \begin{equation}
   E_{f}(z)\ =\ {\exp}\{ -S_{\Upsilon}(z)-S_{B_{c}}(z) \}
   \label{sudakov-f},
   \end{equation}
   \begin{equation}
   E_{n}(z)\ =\ {\exp}\{ -S_{\Upsilon}(z)-S_{B_{c}}(z)-S_{V}(z) \}
   \label{sudakov-n},
   \end{equation}
   \begin{equation}
   S_{\Upsilon}(z)\ =\
   s(x_{1},p_{1}^{+},1/b_{1})
  +2{\int}_{1/b_{1}}^{z}\frac{d{\mu}}{\mu}{\gamma}_{q}
   \label{sudakov-bb},
   \end{equation}
   \begin{equation}
   S_{B_{c}}(z)\ =\
   s(x_{2},p_{2}^{+},1/b_{2})
  +2{\int}_{1/b_{2}}^{z}\frac{d{\mu}}{\mu}{\gamma}_{q}
   \label{sudakov-bc},
   \end{equation}
   \begin{equation}
   S_{V}(z)\ =\
   s(x_{3},p_{3}^{+},1/b_{3})
  +s(\bar{x}_{3},p_{3}^{+},1/b_{3})
  +2{\int}_{1/b_{3}}^{z}\frac{d{\mu}}{\mu}{\gamma}_{q}
   \label{sudakov-ds},
   \end{equation}
   \begin{equation}
   H_{f}({\alpha},{\beta},b_{i},b_{j}) \ =\
   K_{0}(b_{i}\sqrt{-{\alpha}})
   \Big\{ {\theta}(b_{i}-b_{j})
   K_{0}(b_{i}\sqrt{-{\beta}})
   I_{0}(b_{j}\sqrt{-{\beta}})
   + (b_{i}{\leftrightarrow}b_{j}) \Big\}
   \label{hab},
   \end{equation}
   \begin{eqnarray}
   H_{n}({\alpha},{\beta},b_{2},b_{3}) &=&
   \Big\{ {\theta}(-{\beta}) K_{0}(b_{3}\sqrt{-{\beta}})
  +\frac{{\pi}}{2} {\theta}({\beta}) \Big[
   iJ_{0}(b_{3}\sqrt{{\beta}})
   -Y_{0}(b_{3}\sqrt{{\beta}}) \Big] \Big\}
   \nonumber \\ &{\times}&
   \Big\{ {\theta}(b_{2}-b_{3})
   K_{0}(b_{2}\sqrt{-{\alpha}})
   I_{0}(b_{3}\sqrt{-{\alpha}})
   + (b_{2}{\leftrightarrow}b_{3}) \Big\}
   \label{hcd},
   \end{eqnarray}
  where the form of $s(x,Q,1/b)$ can be found in Ref.\cite{pqcd1};
  ${\gamma}_{q}$ $=$ $-{\alpha}_{s}/{\pi}$ is the quark
  anomalous dimension;
  $I_{0}$, $J_{0}$, $K_{0}$ and $Y_{0}$ are Bessel functions;
  the gluon virtuality ${\alpha}_{g}$,
  the quark virtuality ${\beta}_{i}$, and scale $t_{i}$
  are defined as follows.
   \begin{eqnarray}
  {\alpha}_{g} &=& \bar{x}_{1}^{2}m_{1}^{2}
                +  \bar{x}_{2}^{2}m_{2}^{2}
                -  \bar{x}_{1}\bar{x}_{2}t
   \label{gluon-q2-e}, \\
  {\beta}_{a} &=& m_{1}^{2} - m_{b}^{2}
               +  \bar{x}_{2}^{2}m_{2}^{2}
               -  \bar{x}_{2}t
   \label{beta-fa}, \\
  {\beta}_{b} &=& m_{2}^{2} - m_{c}^{2}
               +  \bar{x}_{1}^{2}m_{1}^{2}
               -  \bar{x}_{1}t
   \label{beta-fb}, \\
  {\beta}_{c} &=& x_{1}^{2}m_{1}^{2}
               +  x_{2}^{2}m_{2}^{2}
               +  \bar{x}_{3}^{2}m_{3}^{2}
   \nonumber \\ &-&
                  x_{1}x_{2}t
               -  x_{1}\bar{x}_{3}u
               +  x_{2}\bar{x}_{3}s
   \label{beta-fc}, \\
  {\beta}_{d} &=& x_{1}^{2}m_{1}^{2}
               +  x_{2}^{2}m_{2}^{2}
               +  x_{3}^{2}m_{3}^{2}
    \nonumber \\ &-&
                  x_{1}x_{2}t
               -  x_{1}x_{3}u
               +  x_{2}x_{3}s
   \label{beta-fd}, \\
   t_{a(b)} &=& {\max}(\sqrt{-{\alpha}_{g}},\sqrt{-{\beta}_{a(b)}},1/b_{1},1/b_{2})
   \label{tab}, \\
   t_{c(d)} &=& {\max}(\sqrt{-{\alpha}_{g}},\sqrt{{\vert}{\beta}_{c(d)}{\vert}},1/b_{2},1/b_{3})
   \label{tcd}.
   \end{eqnarray}

  \end{appendix}

  
  \end{document}